\documentstyle[12pt,epsfig]{article}
\begin{document}
\def\pio{$\pi ^o$}
\def \mpla#1#2#3{Mod. Phys. Lett. A {\bf#1}, #2 (#3)}
\def \nc#1#2#3{Nuovo Cim. {\bf#1}, #2 (#3)}
\def \np#1#2#3{Nucl. Phys. {\bf#1}, #2 (#3)}
\def \pisma#1#2#3#4{Pis'ma Zh. Eksp. Teor. Fiz. {\bf#1}, #2 (#3) [JETP Lett.
{\bf#1}, #4 (#3)]}
\def \pl#1#2#3{Phys. Lett. {\bf#1}, #2 (#3)}
\def \plb#1#2#3{Phys. Lett. B {\bf#1}, #2 (#3)}
\def \nim#1#2#3{Nucl. Instr.\& Meth. {\bf#1}, #2 (#3)}
\def \pr#1#2#3{Phys. Rev. {\bf#1}, #2 (#3)}
\def \prd#1#2#3{Phys. Rev. D {\bf#1}, #2 (#3)}
\def \prl#1#2#3{Phys. Rev. Lett. {\bf#1}, #2 (#3)}
\def \prp#1#2#3{Phys. Rep. {\bf#1}, #2 (#3)}
\def \ptp#1#2#3{Prog. Theor. Phys. {\bf#1}, #2 (#3)}
\def \rmp#1#2#3{Rev. Mod. Phys. {\bf#1}, #2 (#3)}
\def \rp#1{~~~~~\ldots\ldots{\rm rp~}{#1}~~~~~}
\def \yaf#1#2#3#4{Yad. Fiz. {\bf#1}, #2 (#3) [Sov. J. Nucl. Phys. {\bf #1},
#4 (#3)]}
\def \zhetf#1#2#3#4#5#6{Zh. Eksp. Teor. Fiz. {\bf #1}, #2 (#3) [Sov. Phys. -
JETP {\bf #4}, #5 (#6)]}
\def \zpc#1#2#3{Zeit. Phys. C {\bf#1}, #2 (#3)}
\def\etal{et al.}
\def\dec{\rightarrow}
\def\ups{$\Upsilon (4{\rm S})$}
\def\bar{\overline}
\def\B{$B$}
\def\BB{$\overline{B}$}
\def\etap{$\eta ^{\prime}$}
\def\bbar{$B\bar{B}$}
\pagenumbering{arabic}
\title{ 
GETTING READY FOR CP VIOLATION IN $B$ DECAYS\footnote{Talk given at Les Rencontres
de Physique le La Vall\`ee d' Aoste, La Thuile, March 1-6, 1999.}
\author{
Marina Artuso       \\
{\em Department of Physics} \\
Syracuse University        \\
Syracuse, NY 13244
\\
\vspace{1cm} 
(CLEO Collaboration)}}
\maketitle
\baselineskip=14.5pt
\begin{abstract}
We review some recent results on b decays relevant to future CP violation
studies. In particular, recent data on 
rare hadronic $B$ decays will be presented, as well as inclusive searches for evidence
for gluonic penguins and   first hints of final state interaction phases in $B\dec D^{\star}\rho$. 
The results discussed are based on data collected 
by the CLEO experiment including up to $5.8\times 10 ^6$ \bbar\ pairs.
\end{abstract}
\baselineskip=17pt
\newpage
\section{Introduction}
In the framework of the Standard Model the gauge bosons, $W^{\pm}$, 
$\gamma$ and 
$Z^o$ couple to  
mixtures of the physical $d,~ s$ and $b$ states. This mixing is described
by the Cabibbo-Kobayashi-Maskawa (CKM) matrix:
\begin{equation}
V_{CKM} =\left(\begin{array}{ccc} 
V_{ud} &  V_{us} & V_{ub} \\
V_{cd} &  V_{cs} & V_{cb} \\
V_{td} &  V_{ts} & V_{tb}  \end{array}\right).
\end{equation}
A commonly used approximate parameterization was originally proposed by 
Wolfenstein \cite{wolf}. 
It reflects the hierarchy between the magnitude of matrix elements 
belonging to different 
diagonals. It is defined as:
\begin{equation}
V_{CKM} =\left(\begin{array}{ccc} 
1-\lambda ^2/2&  \lambda &  A\lambda ^3(\rho -i\eta(1-\lambda ^2/2)\\
-\lambda & 1-\lambda ^2/2-i\eta A^2\lambda ^4 & A\lambda ^2 (1+i\eta\lambda ^2)\\
A\lambda ^3(1- \rho -i\eta)&  -A\lambda ^2& 
1\end{array}\right).\end{equation}

The Standard Model parameterization of the quark mixing via the CKM 
matrix element 
accomodates a complex phase, and therefore offers a 
natural way to model the intriguing phenomenon of $CP$ violation. So far this 
violation has 
been measured only in neutral $K$ decays. 

\begin{figure}[hbt]
\centerline{\epsfig{figure=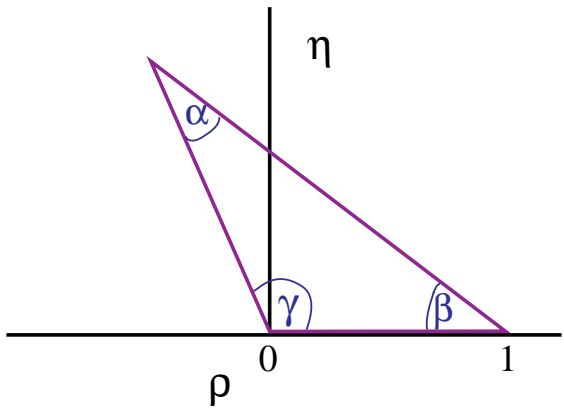, width=3in}}
 \caption{\it
      Unitarity Triangle for CKM elements.
   \label{ckm_tri} }
\end{figure} 

The CKM matrix must be unitary and the relation between elements of different
rows dictated by this property can be graphically represented as
so called `unitarity triangles'.   Fig.~\ref{ckm_tri} shows one of the
triangles where all the angles are expected to be large: 
the angles $\alpha$, $\beta$ and 
$\gamma$ are all 
related to the single phase in the $CKM$ matrix element.
The study of $B$
decays will eventually
allow the 
measurements of  all the three angles. Additional constraints on the sides
will be available too, through more precise measurements of $V_{ub}/V_{cb}$ and 
the determination of the 
$B^o_s\bar{B}^o_s$ mixing parameter $x_s$.  

In parallel, the study of rare decays can provide a window beyond the Standard 
Model, through a detailed comparison of measured and expected branching fractions.
 This study needs a refined understanding of strong
interaction effects, but may provide powerful constraints on a 
wide spectrum of models that try to address some of the shortcomings of the
Standard Model. 

\section{Rare Decays}
Rare $b$ decays encompass several different final states. In general their 
common feature is that their dominant decay diagram is based on a suppressed 
mechanism, either because it is a higher order term in a series expansion
(e.g. loops in the so called `penguin' diagrams) or because the quark coupling
at the decay vertex is CKM suppressed.

\begin{figure}[htb]
\vspace{1.0cm}
\centerline{\epsfig{figure=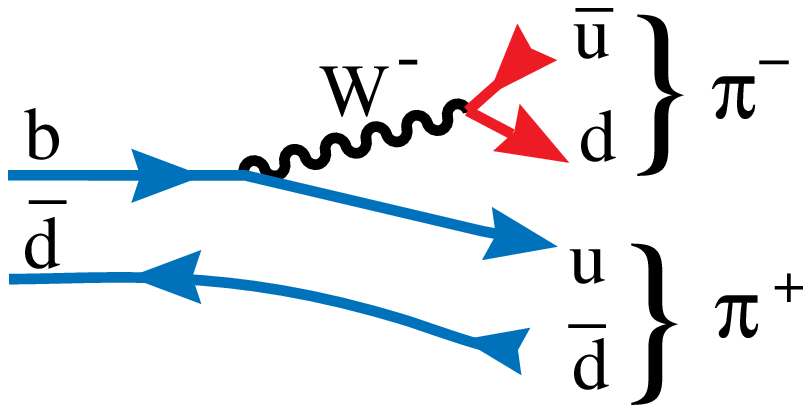, width=2.7in}
\epsfig{figure=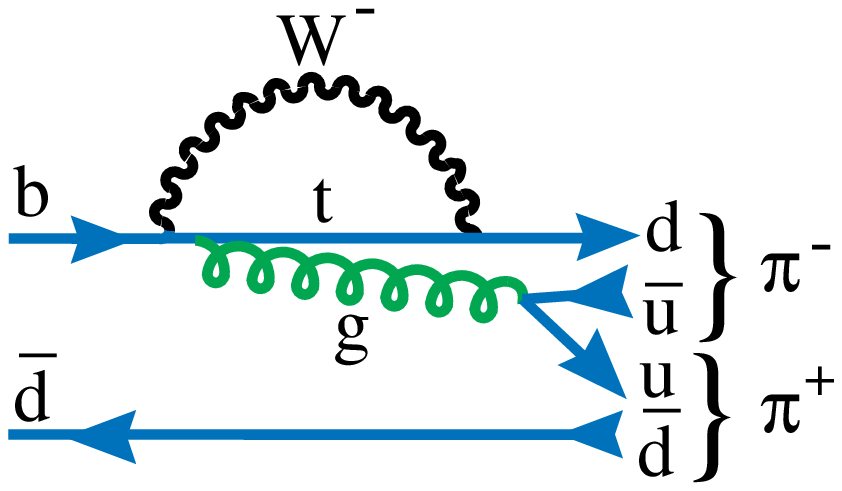, width=2.7 in}}
 \caption{\it
      Feynman Diagrams for the decay 
      $B^0\dec \pi ^+\pi ^-$.
   \label{p-poll} }
\end{figure} 

There are several reasons why a precise experimental mapping of the 
phenomenology of rare $b$ decays is very important. 
First of all the decay amplitude suppression makes it possible for these 
Standard Model processes to interfere with decay diagrams mediated by exotic 
mechanisms due to `beyond Standard Model' interactions. In addition, loop 
diagrams and CKM suppression can affect our ability of measuring CP violation 
phases in two different ways. On one hand, loops and CKM suppressed diagrams 
can lead to final states accessible to both $B$ and $\bar{B}$ decays, 
making it possible to measure interference effects even without neutral
$B$ mixing. On the
other hand, the interplay between these two processes can cloud the 
relationship between measured $CP$ asymmetries and the CKM phase when mixing
induced CP violation is looked for. A classical 
example 
of this effect is the decay $B^o\dec \pi^+\pi^-$. The two Standard Model 
diagrams contributing to this decay process are shown in Fig.~\ref{p-poll}. 
If the $b\dec u$ diagram is dominant, the angle $\alpha$ can be extracted 
from the measurement of the asymmetry in the decay $B^o\dec \pi^+
\pi ^-$.  On the other hand, if these two diagrams 
have comparable amplitude, the extraction of $\alpha$ from this decay channel is
 going to be a much more difficult task. 

CLEO has studied several decays that can lead to a more precise understanding 
of the interplay between penguin diagrams and $b\dec u$ diagrams in $B$ meson
decays. The analysis technique used has been extensively refined in order to 
make the best use of the limited statistics presently available. In most of the
decay channels of interest, the dominant source of background are continuum
events; the fundamental difference
between $e^+e^-\dec q\bar{q}$ and $B$ decays is the shape of the underlying event. The
latter decays tend to produce a more `spherical' distribution of particles
whereas continuum events tend to be more `jet-like', with most of the particle
emitted into two narrow back to back `jets'. This property can be translated 
into
several different shape variables.
 CLEO constructs a Fisher discriminant 
${\cal F}=\Sigma _i \alpha_ iy_i$, a linear combination of several 
variables $y_i$. The variables used are $|\cos{\theta _{cand}}|$, the
cosine of the angle between the candidate $B$ sphericity axis and the beam
axis, the ratio of Fox-Wolfram moments $R_2=H_2/H_0$, and nine
variables that measure the scalar sum of the momenta of the tracks and showers
from the rest of the event in 9 angular bins, each of 10$^{\circ}$,
around the candidate sphericity axis. The coefficients $\alpha _i$ have
been chosen to optimize the separation between signal and background Monte 
Carlo samples \cite{rareb}. In 
addition, several kinematical constraints allow a more precise determination of
the final state. First of all  the energy difference $\Delta E=E_{cand}-E_B$, where $E_{cand}$ is
the reconstructed  candidate mass  and $E_B$ is the known beam 
energy ($\Delta E =0$ for signal events) and the
beam constrained mass $M$. In addition, the $B$ decay angle with 
respect to the beam axis has a $\sin ^2 (\theta _B )$ angular distribution. 
Finally,
to improve the separation between the final states $K\pi$ and $\pi\pi$ 
the specific energy loss in the drift chamber, $dE/dx$ is used. 

CLEO uses a
sophisticated unbinned maximum likelihood (ML) fit to optimize the precision of the
signal yield obtained in the analysis, using 
$\Delta E$, $M$, ${\cal F}$, $|\cos{\theta _B}|$, and $dE/dx$ wherever 
applicable . In each of these fits the likelihood of the event is parameterized
by the sum of probabilities for all the relevant signal and background 
hypotheses, with relevant weights determined by maximizing the likelihood 
function (${\cal L}$). The probability of a particular hypothesis is calculated
as the product of the probability density functions for each of the input 
variables determined with high statistics Monte Carlo samples.
The likelihood function is defined as:
\begin{equation}
{\cal L} = \Pi _k \Sigma _i P^i_k(\Delta E, M_B,{\cal F}, dE/dx, 
\cos{\theta _B}) \times f_i
\end{equation}
where the index $k$ runs over the number of events, the index $i$ over
the hypotheses considered, $P^i_k$ are the probabilities for
different hypotheses obtained from Monte Carlo
simulations of the signal and background channels considered and 
independent data samples, and $f_i$ are 
the fractional yields for 
 hypothesis $i$, with the constraint:
\begin{equation}
\Sigma _i f_i =1
\end{equation}

Further details about the likelihood fit can be found elsewhere \cite{rareb}.
 The fits
include all the decay channels having a similar topology. For example, in the
final state including two charged hadrons, the final states considered were
$K^{\pm}\pi^{\mp}$, $\pi^{\pm}\pi^{\mp}$ and $K^{\pm}K^{\mp}$.

\begin{figure}
\vspace{-2.0cm}
\centerline{\epsfig{figure=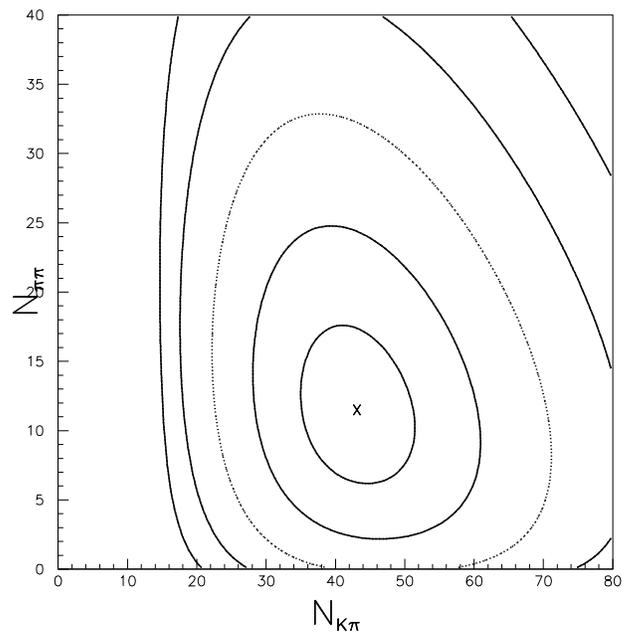, width=3.8in}}
\caption{\it  $n\sigma$ contours for $B^o\dec h^+\pi ^-$, corresponding 
to an increase of $-2\ln{\cal L}$ by $n^2$. The cross identifies the most
probable values of the $\pi^+pi^-$ and $K^+\pi^-$ yields.   
   \label{pipi} }
\end{figure}

Fig. \ref{pipi} shows contour plots of the ML fits for the signal yields
in the  $K^{\pm}\pi^{\mp}$ and $\pi^{\pm}\pi^{\mp}$ final states. The other
channels included in the likelihood function have $f_i$ fixed to their most
probable value extracted from the fit. It can 
be seen that there is a well defined signal for the   
$K^{\pm}\pi^{\mp}$ final state, whereas
there is less than 3$\sigma$ evidence of having seen 
$\pi^+\pi^-$. This shows that the $b\dec u$ diagram
is suppressed with respect to the penguin diagram in $B$ decays to two 
pseudoscalar mesons. Table \ref{twops} summarizes the CLEO results for the 
$B\dec K\pi,~\pi \pi,~ KK$ final states. Unless explicitly stated, the 
results are based 
on a data sample of 5.8 million $B\bar{B}$ pairs. There is a consistent 
pattern of penguin dominance in \B\ decays into two charmless 
pseudoscalar
mesons that makes the prospects of extracting the angle  
$\alpha$ from the study of the CP asymmetry in
the $B\dec \pi\pi$ mode less favorable than originally expected.

\begin{figure}[htb]
\centerline{\epsfig{figure=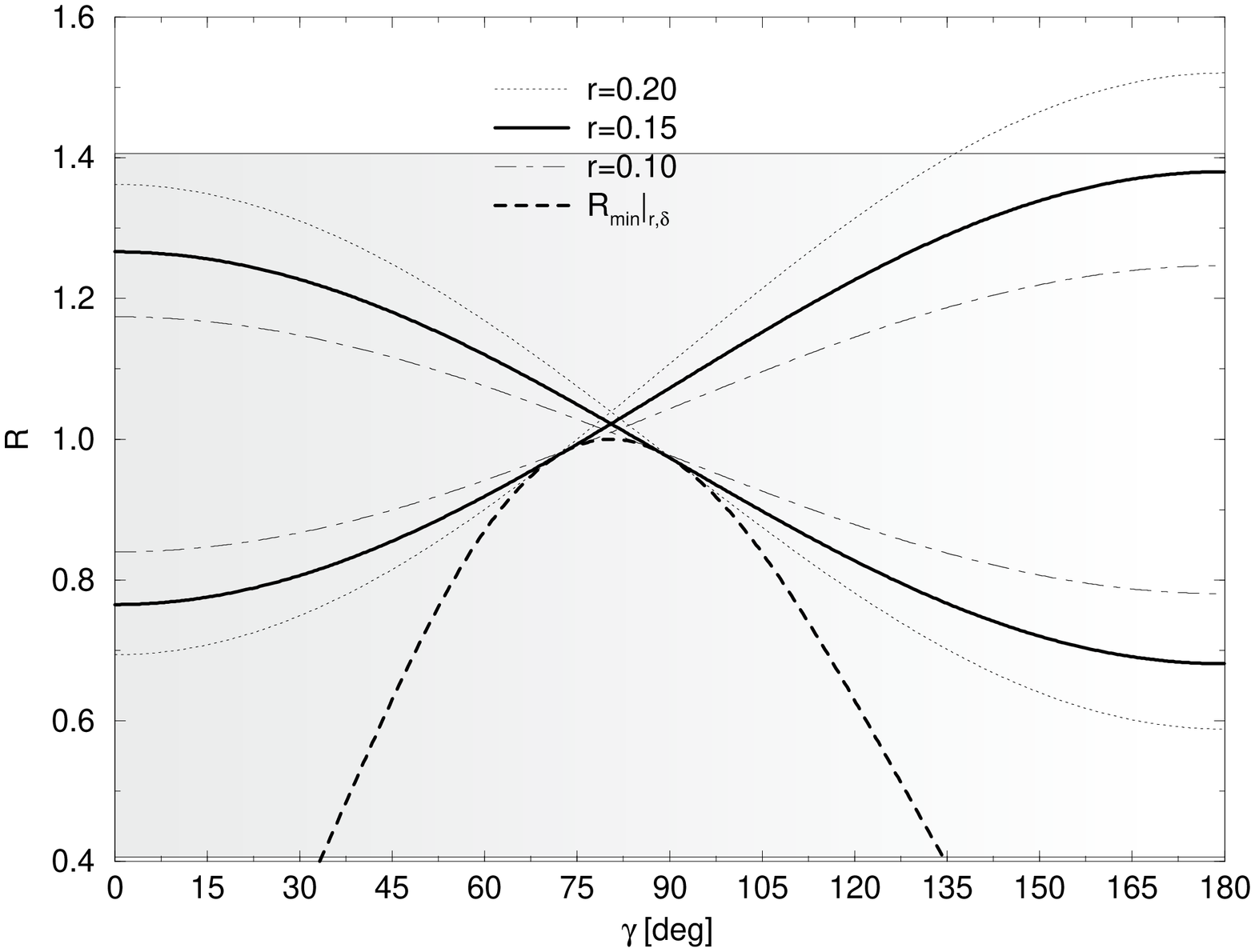, width=3.8in}}
\caption{\it The dependence of the extremal values of $R_{FM}$ on
the CKM angle $\gamma$ in the case of negligible rescattering 
effects \cite{buras}; $r = \mid T \mid/\sqrt{<\mid P\mid ^2>}$, where $T$ is
the
``tree'' amplitude and $P\equiv {\cal A}(B^+\dec \pi ^+ K^o)$, and 
$R_{min}\mid_{r\delta}$ is the minimal  value of $R_{FM}$.
\label{f-buras} }
\end{figure}

Recently a lot of theoretical discussion has been focused on the possibility
of extracting the angle $\gamma$ from the study of a variety of ratios of the
decays just reviewed. Originally Fleisher and Mannel \cite{fm} proposed to consider the ratio

\begin{equation}
R_{FM}=\frac{{\cal B}(B^o\dec K^+\pi ^-)}{{\cal B}(B^+\dec K^o\pi ^+)}
\label{fm}
\end{equation}
Subsequently Neubert and Rosner \cite{nr} proposed a different ratio:
\begin{equation}
R_{NR}=\frac{{\cal B}(B^+\dec K^o\pi ^+)}{2{\cal B}(B^+\dec K^+\pi ^o)}
\label{nr}
\end{equation}

A careful study of the effects of final state interaction is necessary 
to evaluate the relationship between these two ratios 
and $\gamma$ \cite{buras}, \cite{neubert}. The CLEO result for these ratios is:

\begin{eqnarray}
R_{FM} & = & 0.9 \pm 0.4\pm 0.2 \pm 0.2 \\
R_{NR} & = & 0.47\pm 0.21 \pm 0.12
\end{eqnarray}
The last error in $R_{FM}$ is due to the uncertainty in the charged to neutral
$B$ meson ratio at the \ups\ ($f_{\pm}/f_{oo}=1.15\pm 0.17\pm 0.06$) 
\cite{silvia}. The experimental
errors are too big to be able to exclude specific $\gamma$ intervals, but 
Fig. \ref{f-buras} shows that a more precise measurement could restrict
the allowed $\gamma$ regions.

\begin{table}
\centering
\caption{ \it Summary of CLEO results for $B\dec K\pi, \pi\pi, KK$. }
\vskip 0.1cm 
\begin{tabular}{|l|c|c|c|} \hline
Mode          & Yield  &${\cal B}$/U.L. (x10$^{-5}$) & Theory (10$^{-5}$)\cite{ali}\\
\hline
$K^{+}\pi ^{-}$ & $43.1^{+9.0}_{-8.2}$ & $1.4\pm 0.3 \pm 0.2$ & 1.4 -1.8 \\
$K^{+}\pi^{o}$ & $38.1^{+9.7}_{-8.7}$ & $1.5\pm 0.4 \pm 0.3$ & 0.9 - 1.2 \\
$K^{o}\pi ^{-}$ & $12.3^{+4.7}_{-3.9}$ & $1.4\pm 0.5 \pm 0.2$ & 1.4 -2.2 \\
\hline
$\pi^{+}\pi ^{-}$ & $11.5^{+6.3}_{-5.2}$ & $<0.84$ & 0.9 -1.2 \\
$\pi^{+}\pi^{o}$ & $14.9^{+8.1}_{-6.9}$ & $<1.6$ & 0.3 -0.7 \\
\hline
$K^{+}K^{-}$ & $0.0^{+1.6}_{-0.0}$ & $<0.23 $ & 0.3 -0.7 \\ 
$K^{+}K^{o}$ & $1.8^{+2.6}_{-1.4}$ & $<0.93 $ & 0.3 -0.6 \\
\hline
\multicolumn{4}{|c|}{Published Results based on $3.3\times 10^6 B\overline{B}$ pairs}
\\ \hline
$K^{o}\pi ^{o} $ & $4.1^{+3.1}_{-2.4}$ & $<4.1 $ & 0.5 -0.7 \\
$\pi^o\pi^o$ & $2.7^{+2.7}_{-1.7}$ & $<0.93$ & 0.3-0.6 \\
$K^o\bar{K}^o$ & 0 & $<1.7$ &0.1 -0.8\\  
\hline
\end{tabular}
\label{twops}
\end{table}

The CLEO study of $B$ decays to final states including two charmless hadrons
has presented some other interesting surprises. A large branching ratio
has been discovered for final states including a $\eta ^{\prime}$ meson.
The analysis technique in this case is essentially the same as the one discussed
above. The $\eta ^{\prime}$ is reconstructed both in its
$\eta \pi^+\pi^-$ and $\rho \gamma$ decay channels. The results for different decay modes including 
$\eta$ and $\eta ^{\prime}$ are summarized in Table~\ref{etaptab}. The most notable 
feature of these results is the astonishing large rate for the 
$\eta ^{\prime} K$ final state. While several theoretical interpretations have
been proposed to explain this enhancement \cite{bizarre}, a very
plausible explanation is still pointing to a dominance of penguin effects in
$B$ decays into two pseudoscalar charmed hadrons.

\begin{table}
\centering
\caption{ \it Summary of CLEO results for $B\dec \eta ^{\prime}\pi/K,
\eta \pi/K$.
}
\vskip 0.1cm 
\begin{tabular}{|l|c|c|c|} \hline
Mode          & Yield  &${\cal B}$/U.L. (x10$^{-5}$) & Theory (10$^{-5}$)\cite{ali}\\
\hline
$\eta ^{\prime}K^+$ & $68.6$ & $7.4^{+0.8}_{-1.3}\pm 1.0$ & 2.1-3.5 \\
$\eta ^{\prime}K^o$ & $18.1$ & $5.9^{+1.8}_{-1.6} \pm 0.9$ & 2.0-3.5 \\
$\eta ^{\prime}\pi ^{+}$ & 1.0 & $<1.2$ & 1.1-2.7 \\
\hline
$\eta K^{+}$ &1.3 & $<1.4$ & 0.2-0.4 \\ 
$\eta \pi^+$ & 0.2 & 1.5 & 0.2-0.2 \\  
\hline
\end{tabular}
\label{etaptab}
\end{table}

\begin{table}
\centering
\caption{ \it Summary of CLEO results for $\overline{B}^o/B^- \dec \rho\pi/K$ 
(in units of $10^{-5})$}
\vskip 0.1cm 
\begin{tabular}{|l|c|c|c|c|c|} \hline
~~& $\rho ^-\pi^+$ & $\rho ^+\pi^-$ 
& $\rho ^-\pi ^0$ & $ \rho ^o\pi ^-$ & 
$\rho ^o\pi ^o$ \\
\hline
CLEO ${\cal B}$/U.L. ) & \multicolumn{2}{c|}{ $3.5^{+1.1}_{-1.0}\pm 0.5$}
& $<7.7$ & $1.5 \pm 0.5 \pm 0.4$ & $<1.8$ \\ 
\hline
Ciuchini \etal \cite{it} & 1.0-7.5 & 0.2-1.9 &0.3-2.6 & 0.5-1.1 & 0.0-0.2 \\
Ali \etal \cite{ali} & 2.1-3.4 & 0.6-0.9 & 1.1 -1.6 & 0.1-0.7 &0.0-0.2 \\
\hline
\end{tabular}
\label{pv}
\end{table}

\begin{figure}[htb]
\centerline{\epsfig{figure=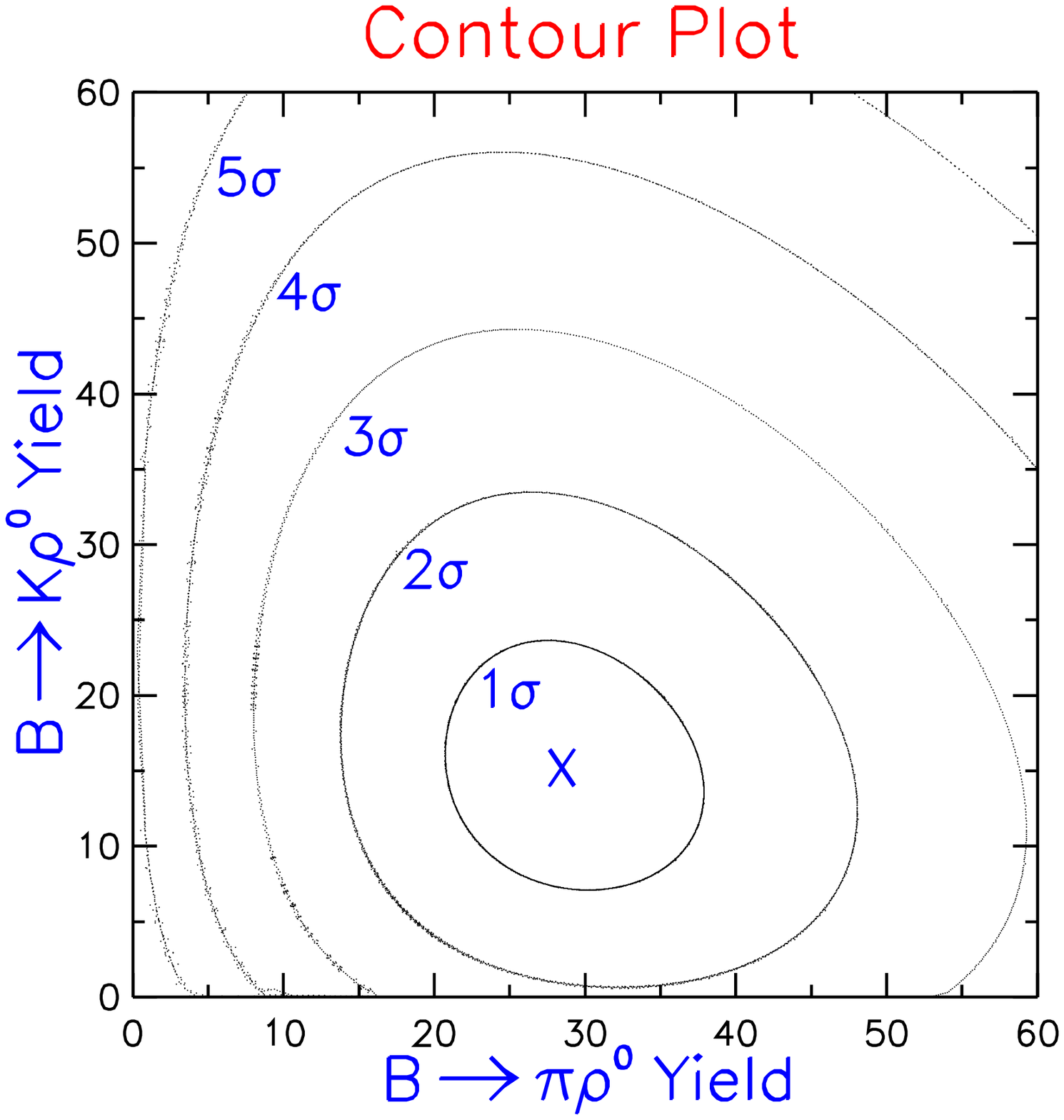, width=3.8 in}}
\caption{\it $n\sigma$ contours for the decays $B^+ \dec \rho ^0 \pi^+$
and $B^+\dec \rho ^0 K^+$. The cross identifies the most probable values of the yields
for these two channels.
\label{pirho}}
\end{figure}

The results discussed so far are quite discouraging for our prospects of 
studying $b\dec u$ transitions in charmless hadronic $B$ decays. Luckily recent
CLEO data suggest that final states involving a vector and a psedoscalar meson 
offer a different picture. In fact,  the first observation of 
$B^+ \dec \pi ^+ \rho^o$ and $B \dec \rho ^{\pm} \pi ^{\mp} $\cite{gao}. 
They are 
also first observations
of $b\dec u$ hadronic transitions. The analysis procedure is similar to the one
adopted for other charmless exclusive decays. In this case, there are three 
particles in the final state and some additional constraints are provided
by the vector particle decay kinematics. The invariant mass of its decay
products must be consistent with the vector meson mass. In addition, the
vector meson is polarized, thus its helicity angle
$\theta _h$ is expected to have a $\cos ^2{\theta _h}$ distribution. In this
 case the maximum likelihood fit
includes $\pi\rho^0$ and $K\rho ^0$ signal channels and continuum samples. The $n\ \sigma$ contour
 plot for this analysis is shown 
in Fig. \ref{pirho} and gives solid evidence for a $B^+ \dec \pi ^+\rho ^o$ 
signal, while in this case the $B^+\dec K^ +$ channel appears to be suppressed.
Other $b\dec u$ modes with $B\dec PV$ topology have been searched: the measured upper limits
are very close to the theoretical predictions, as shown in Table~\ref{pv}. Thus, hopefully, 
more positive signals can be measured soon, when the full CLEO II.5 data
set will be processed.\footnote{The data sample used in this analysis comprises $5.8\times 10^6
B\bar{B}$ pairs, while the full data sample available upon completion of the
CLEO II.5 phase is $ 9 \times 10^6 B\bar{B}$ pairs.}

\section{Inclusive $\eta ^{\prime}$ production}
A deeper understanding of the dynamics of the gluonic penguin process $b\dec sg$ is quite
important as it is expected to play a critical role in direct CP violation in
B decays \cite{cpdirect}. As mentioned above, this process appears to
be the dominant decay mechanism in $B$ decays into two charmless pseudoscalar 
mesons. The inclusive $B\dec \eta ^{\prime} X_s$ signal, where the $X$ represents
a collection of particles containings a single $s$ quark, is a possible signature
of $b\dec sg^{\star}, g^{\star}\dec g u\bar{u}, d\bar{d}\ {\rm or}\ s\bar{s}$. 

\begin{figure}[htb]
\centerline{\epsfig{figure=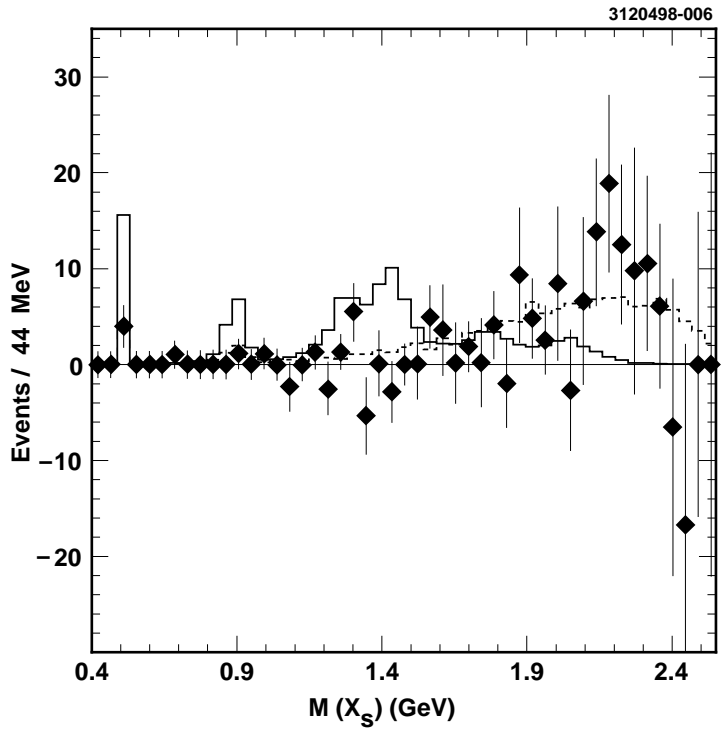, width=3.8 in}}
\caption{\it  The continuum subtracted $M(X_s)$ distribution (points with error bars) with
expected $M(X_s)$ distribution for a mixture of two body $b\dec sq\overline{q}$ (solid) and
three-body $b\dec s g^{\star}$ with $g^{\star}\dec g \eta ^{\prime}$ simulated with JETSET 
(dashed). The data points have been corrected for $M(X_s)$ dependent efficiency. Each 
simulation has been normalized to the data yield.
   \label{mxs} }
\end{figure} 

In searching for $B\dec \eta ^{\prime} X_s$, we focus on the end point of the $\eta ^{\prime}$
spectrum, to suppress contributions from $b\dec c$ processes. In this analysis
the momentum window is chosen to be 
$2.0 GeV/c~ <p_{\eta ^{\prime}} < 2.7~ GeV/c$. The
analysis technique is similar to the one that lead to a successful measurement
of the inclusive $b\dec s\gamma$ process, explicitly reconstructing 
the $X_s\dec \eta ^{\prime} K^{\pm}n\pi$,
where $n\le 4$ and at most 1 \pio\ is included. This pseudo-exclusive reconstruction technique is
quite effective in suppressing continuum background. Upon continuum subtraction an excess 
yield is found. Fig. \ref{mxs} shows the $X_s$ mass spectrum compared with different Monte
Carlo shapes obtained with two different hypotheses for the source of this excess: $b\dec sq\bar{q}$
and $b\dec sg^{\star}$. The mass spectrum clearly favors a $b\dec sg^{\star}$ origin.
Several alternative hypotheses have been investigated
as an explanation of this excess \cite{etapconf}, but all the features of the data seem to
favor the $b\dec sg^{\star},~g^{\star}\dec g\eta ^{\prime}$ hypothesis. 

The branching fraction corresponding to this excess is:
\begin{equation}
{\cal B}(B\dec \eta^{\prime}X_s) =(6.2\pm 1.6 \pm 1.3 ^{+0.0}_{-1.5})\times 10^{-4}\ (2.0~ GeV/c 
<p_{\eta^{\prime}}
<2.7~ GeV/c)
\end{equation}

This measured \etap\ production is probably too large compared to conventional
calculations of the hadronic matrix element for the penguin 
operator \cite{datta}. Explanations
for this anomaly are similar to the ones proposed for the enhancement in exclusive
modes discussed above \cite{etapexcl}. A more precise theoretical evaluation of the 
hadronic matrix element is necessary before we can advocate new physics to understand this excess.

\section{Final state interaction phases}
Different final state interaction phases are a necessary ingredient, as well as different weak 
interaction phases, to lead to the interference that produces direct CP asymmetries. A
way to identify final state interaction effects is to search for phase differences in decays
into two vector states. This information is obtained by performing a full angular distribution
analysis of such decays. 

The formalism and analysis technique will be discussed with reference to the decay $B\dec D^{\star} \rho$, 
the mode studied most recently \cite{dsr}. While the charged $B$ decay is mediated both
by a spectator and color suppressed diagram, only the spectator diagram is expected to
contribute to the neutral \B\ decay. Thus interference effects in the latter decay are an unambiguous
sign of final state interaction.

The angular distribution for $B\dec D^{\star}\rho$ 
is described in terms of three helicity amplitudes, $H_{\pm,o}$:

\begin{eqnarray}
\lefteqn{\frac{32\pi}{9\Gamma}\frac{d^3\Gamma}{d\cos{\theta _1}d\cos{\theta _2}d\chi} = } \nonumber\\
& & 4 \mid H_o\mid ^2 \cos ^2{\theta_ 1}\cos ^2{\theta _2} + (\mid H_{+}\mid ^2 + \mid H_\mid ^2)
\sin ^2 {\theta _1}\sin ^2 {\theta _2}  + \nonumber\\
& & [{\cal R}(H_-H^{\star}_+)\cos{2\chi} + {\cal I}(H_-H^{\star}_+)\sin{2\chi}]2\sin ^2 {\theta _1}
 \sin ^2 {\theta _2}  +  \nonumber   \\  
& & [{\cal R}(H_-H^{\star}_o-H_+H^{\star}_o)\cos{\chi} + 
{\cal I}(H_-H^{\star}_o -H_-H^{\star}_o)\sin{\chi}] \sin{2\theta _1}\sin{2\theta _2}
\sin ^2 {\theta _2}, 
\label{angdist}
\end{eqnarray}

where $\theta _1$ is the $D^0$ angle in the $D^{\star}$ rest frame, $\theta _2$ is the $\pi ^+$
angle in the $\rho$ rest frame and $\chi$ is the angle between  the $D^{\star}$
and the $\rho$ decay planes in the $B$ rest frame. Assuming $CP$ symmetry in the decay, the
helicity amplitudes of the \B\ and \BB\ are related:

\begin{equation}
H_{\lambda }(\bar{B}) = H_{-\lambda }(B)\ (\lambda =+,-,0). 
\end{equation}

This corresponds to flipping the sign of $\chi$. The two data sets are combined accordingly. 
Furthermore, the phases are measured with respect to $H_o$ and the normalization condition is:
\begin{equation}
\mid H_o\mid ^2 + \mid H_+\mid ^2 +\mid H_-\mid ^2 =1
\end{equation}

This makes the angular distribution invariant under the exchange $H _{\pm} \leftrightarrow H^{\star }_{\mp}$,
thus making the distinction between $H_+$ and $H_-$ arbitrary. Because of
the $V-A$ nature of the interaction, $H_-$ is assumed to be the larger of the two 
for \BB . Table \ref{dstarrho} shows the fit results. There are signs of non-zero 
phases. The angular distribution in Eq. \ref{angdist} would imply that non-trivial phases
should result in asymmetric $\chi$ angular distributions. The data do not show
such an asymmetry, probably because of limited statistics. The measured non-zero phases 
are an interesting indication that another ingredient necessary to see CP violation, final
state interaction, indeed exists.

\begin{table}
\centering
\caption{ \it Fit results for helicity amplitudes. The phase and magnitude of the amplitudes
$H_+$ and $H_-$ are measured with respect to $H_o$. 
\label{dstarrho}}
\vskip 0.1cm 
\begin{tabular}{|l|c|c|} \hline
$B^o\dec D^{\star -}\rho ^+$ & Magnitude & Phase \\
\hline
$H_o$ & 0.936 & 0 \\
$H_-$ & $0.317\pm 0.052 \pm 0.013$ & $0.19\pm 0.23 \pm 0.14$ \\
$H_+$ & $0.152\pm 0.058 \pm 0.037$ & $1.47\pm 0.37 \pm 0.32$ \\
\hline
$B^+\dec D^{\star o}\rho ^+$ & Magnitude & Phase \\
\hline
$H_o$ & 0.932 & 0 \\
$H_-$ & $0.283\pm 0.068 \pm 0.039$ & $1.13\pm 0.27 \pm 0.17$ \\
$H_+$ & $0.228\pm 0.069 \pm 0.036$ & $0.95\pm 0.31 \pm 0.19$ \\
\hline
\end{tabular}
\end{table}

\section{Conclusion}
We have reviewed several results on hadronic $B$ decays that are closely
related to the search for direct and mixing mediated CP violation.
In particular, the role of gluonic penguins and final state interaction in these
decays has been investigated extensively. These measurements represent an 
important step in our goal of a precise determination of the CKM parameters.

\section{Acknowledgements}
I would like to thank Mario Greco, Giorgio Bellettini and Giorgio Chiarelli for a very pleasurable
and thought provoking conference. Many thanks to all my colleagues in CLEO that
have made it possible for me to present this wide spectrum of interesting data.
Lastly I would like to thank Sheldon Stone for several interesting 
discussions, Frank W\"urthwein and Tomasz Skwarnicki for useful 
suggestions. This work was supported by NSF.


%
\end{document}